\documentclass[pra,twocolumn,showpacs,superscriptaddress,floatfix]{revtex4-1}
\usepackage[caption=false]{subfig}
\usepackage{graphicx}
\graphicspath{ {} }
\usepackage{amsmath,graphicx}
\usepackage{epsfig}
\usepackage{amsmath,amssymb,mathrsfs,esint}
\usepackage{gensymb}
\usepackage[bookmarks=true,
   colorlinks=true,
   linkcolor=blue,
   urlcolor=blue,
   citecolor=blue,
   bookmarks=true,
   hyperindex=true
]{hyperref}
\usepackage{natbib}
\usepackage{relsize}
\usepackage{float}
\usepackage{dsfont}

\usepackage{soul}

\newcommand{\be}{\begin{equation}}
\newcommand{\ee}{\end{equation}}
\newcommand{\norm}[1]{\left\lVert#1\right\rVert}

\begin{document}

\title{Quasi-conserved charges in the perturbed spin-1/2 XXX model} 

\author{Denis V. Kurlov}
\email[]{d.kurlov@rqc.ru}
\affiliation{Russian Quantum Center, Skolkovo, Moscow 143025, Russia}

\author{Savvas Malikis}
\affiliation{Instituut-Lorentz, Universiteit Leiden, Leiden, The Netherlands}

\author{Vladimir Gritsev}
\affiliation{Institute for Theoretical Physics Amsterdam, Universiteit van Amsterdam, Amsterdam, The Netherlands}
\affiliation{Russian Quantum Center, Skolkovo, Moscow 143025, Russia}

\begin{abstract}
We consider the isotropic spin-$1/2$ Heisenberg spin chain weakly perturbed by a local translationally- and $SU(2)$-invariant perturbation. Starting from the local integrals of motion of the unperturbed model, we modify them in order to obtain quasi-conserved integrals of motion (charges) for the perturbed model. Such quasi-conserved quantities are believed to be responsible for the existence of the prethermalization phase at intermediate timescales. 
We find that for a sufficiently local perturbation the quasi-conserved quantities indeed exist, and we construct an explicit form for the first few of them.
\end{abstract}

\maketitle 

\section{Introduction}

In classical mechanics there is a well understood distinction between integrable and non-integrable systems, as well as between their long-time dynamics. Namely, a generic non-integrable system typically exhibits an ergodic behavior, leading to a chaos,
 whereas integrable systems are non-ergodic and their phase space trajectories are confined to some subregions of the phase space due to the existence of many conserved quantities. 
Moreover, there is a result of tremendous importance, the Kolmogorov-Arnold-Moser (KAM) theorem, which ensures that classical integrable systems under a weak integrability-breaking perturbation are stable for a sufficiently long time \cite{arnold1,arnold2,kolmogorov1,moser1,moser2}. 

Extending the KAM theorem to the quantum case is a long-standing problem. Although recent findings demonstrate some progress in this direction \cite{quantumkam}, a complete understanding is missing and there are numerous open questions.
In part this is due to the fact that in the quantum case even the very definition of integrability is subtle~\cite{quantumint}.

A widely accepted criterion for quantum integrability is that a Hamiltonian $H_0$ is integrable, if there exists a large number of extensive, functionally-independent, and mutually commuting conserved quantities (charges):
\be \label{local_charges}
	\left[ H_0, {\cal Q}_j \right] = \left[ {\cal Q}_k, {\cal Q}_j \right] = 0.
\ee
Importantly, the conserved charges ${\cal Q}_j$ are assumed to be local, in a sense that they are given by the sums of operators with a finite support.

Just like in the classical case, the long time dynamics is very different for integrable and non-integrable systems (to be precise, we do not consider here systems that exhibit Anderson or many-body localization).
Non-integrable systems thermalize according to the Eigenstate Thermalization Hypothesis (ETH), which (at least according to one of the interpretations) means that the total isolated system acts as a heath bath for its own subsystems. This leads to the spread of entanglement over the whole system, such that in the long time limit it is impossible to retrieve any information about the initial state using only local measurements \cite{srendicki, deutschfirst}.
On the contrary, in integrable systems, thermalization is very different. It is described by the generalized Gibbs ensemble (GGE), which takes into account that there are many other conserved quantities apart from the total energy and the number of particles, as it is for the standard grand-canonical ensemble \cite{EF-rev,EF1,poz1}. Moreover, it has been shown that to accurately describe thermalization of integrable systems one should extend the GGE by including not only the local conserved charges, as in Eq. (\ref{local_charges}), but also the {\it quasi-local} ones \cite{quasilocal1,quasilocal2,quasilocalgge,quasilocalgge2}.

Then, in the spirit of the KAM theorem, one may ask what will happen if a quantum integrable system is slightly perturbed away from integrability. What kind of thermalization will it exhibit?
Naively, one would expect that nearly-integrable quantum systems simply thermalize following the ETH.
However, it is widely believed that such systems also exhibit a different, the so-called prethermal, behavior at intermediate times~\cite{prethermalization, prethernearin1, prethermnearin2}. Different studies \cite{pretherml1,pretherml2,pretherml3} suggest that the eventual thermalization occurs at much later times~$t_{\text{th}}\sim \lambda^{-2}$, where~$\lambda \ll 1$ is the strength of the perturbation, and the scaling can be understood from the Golden Rule. Moreover, it is believed that this prethermal phase should be described by some effective GGE \cite{prethgge}. 
Therefore, it is natural to ask what are the charges that define this effective GGE in the prethermal phase. Clearly, since this effective description is only valid at times~$t \lesssim \lambda^{-2}$, these charges can only be {\it quasi}-conserved with the accuracy~${\cal O}(\lambda^2)$. 
In other words, since the exact conservation laws of the unperturbed system constrain the dynamics of an integrable system, one can expect that the dynamics of a perturbed system 
should be restricted by the quasi-conserved charges. 

This idea is also supported by the developments in the context of the slowest operators~\cite{slow1,slow2}. Indeed, for an operator~$O$ that commutes with a Hamiltonian~$H$, the time evolution~$e^{iHt}Oe^{-iHt}$ is trivial. In terms of the quantum information language this means that the information encoded in $O(0)$ does not spread. On the contrary, if~$[O(0),H]\neq 0$, the typical timescale of information spreading is inversely proportional to the norm of~$[O(0), H]$, as follows from the Baker-Campbell-Hausdorff formula. Thus, to slow down the spread of the quantum information one needs to suppress (at least) the first order term in the Baker-Campbell-Hausdorff expansion.

Let us mention that the search for quasi-conserved charges can be linked with an old problem in functional analysis, related to almost commuting matrices \cite{math1} and
explicitly stated by Halmos in \cite{Halmos}.
This long-standing question ``when two almost commuting matrices are close to matrices that exactly commute" was answered eventually by H.~Lin~\cite{Lin}.
More precisely, Lin showed that given~$\epsilon>0$ there exists $\delta>0$ such that if $N\times N$ matrices $A,B$ are Hermitian, with $\parallel AB-BA\parallel<\delta$ and $\parallel A\parallel,\parallel B\parallel\leq1$, then there exists commuting Hermitian $N\times N$ matrices~$X,Y$ such that~$\parallel A-X\parallel+\parallel B-Y\parallel<\epsilon$, where~$\parallel \cdot \parallel$ is a matrix norm. Importantly, $\delta=\delta(\epsilon)$ {\it does not} depend on the dimension $N$. Recently, Hastings obtained an explicit estimate $\epsilon(\delta)\sim \delta^{1/5}$, where the exponent may depend on the choice of the operator norm~\cite{Hastings}. Quite remarkably, the question whether one can find triples of almost commuting matrices has generically a negative answer \cite{triple}. 
A similar story about {\it unitary} matrices is more involved~\cite{math-uni1, note}. There, the existence of almost commuting unitary matrices have some topological obstructions given by the so-called Bott indices. There is an extensive mathematical literature on the subject, see, e.g., Ref.~\cite{math-uni2}.   
Finally, we would like to mention that there is a somewhat related research direction in the context of AdS/CFT correspondence, which deals with the so-called long-range deformed spin chains~\cite{longrange1, longrange2} and $T \bar T$-deformations~\cite{Marchetto2020, Pozsgay2020}. However, these studies deal with the deformations that preserve integrability to all orders in the perturbation strength and thus differ from the present work.

This paper is devoted to the search for quasi-conserved quantities in a quantum spin chain weakly perturbed away from integrability. The rest of the paper is organized as follows. In Section~\ref{S:model_boost} we describe the model and discuss some general properties of the exact conserved charges that are present in the absence of the perturbation. In Section~\ref{S:quasi_charges} we present an ansatz for the quasi-conserved charges and discuss the requirements that the ansatz must satisfy. Finally, in Section~\ref{S:results} we demonstrate our findings, conclude, and formulate some open questions for future research. The paper is supplemented by several technical appendices which clarify our derivations using algebraic tools. Similar techniques can also be used for other spin systems.

\section{The model and conserved charges of the integrable part} 
\label{S:model_boost}

Let us start by describing the specific model that we are going to deal with.
Consider a Hamiltonian 
\be \label{H_tot}
	H_{\lambda} = H_0 + \lambda H_1,
\ee
where $H_0$ is an integrable part, $H_1$ is an integrability-breaking perturbation, and $\lambda > 0$ is a numerical parameter characterizing the perturbation strength. We assume~$\lambda \ll 1$, such that the perturbation is weak.
For the unperturbed system, we take a spin-$1/2$ isotropic Heisenberg spin chain (XXX model) on a one-dimensional lattice:
\be \label{H_0}
	H_0 =  J \sum_{j} \boldsymbol{\sigma}_j\cdot\boldsymbol{\sigma}_{j+1},
\ee
where $\boldsymbol{\sigma}_{j}$ is the vector of Pauli matrices and $J$ is an exchange constant. It is well known that~$H_0$ is integrable, the exact spectrum and the eigenstates can be found using the Bethe ansatz \cite{bethe}, and one has a large number of conserved charges.
Let us now break the integrability by a perturbation of the following form:
\be \label{H_1}
	H_1 =  J \sum_{j}\, \boldsymbol{\sigma}_j\cdot\boldsymbol{\sigma}_{j+2},
\ee
which is nothing other than the next-to-nearest neighbor Heisenberg interaction.
In what follows we put $J = 1$ and assume that the system is in the thermodynamic limit. 
Let us mention that both the unperturbed Hamiltonian~$H_0$ and the perturbation~$H_1$ are translationally- and $SU(2)$-invariant. Also, both are the sums of local operators with the support on two and three sites (for $H_0$ and $H_1$, respectively). Many facts are known about the low-energy properties of this model \cite{WA96}: When $\lambda<\lambda_{c}\approx 0.241$, the model is gapless and is described by a marginally perturbed $SU(2)_{1}$ Wess-Zumino-Witten  model, while for $\lambda_{c}<\lambda\leq 1/2$ the ground state is dimerized (with exactly known Majumdar-Ghosh state at $\lambda=1/2$) with a gap in the spectrum. Spontaneous incommensurate order appears for $\lambda>1/2$. In the limit $\lambda \to \infty$ the Hamiltonian~(\ref{H_tot}) is equivalent to a pair of decoupled XXX models (on even and odd sites), and integrability is restored. However, we do not consider the case of large $\lambda$ and restrict ourselves to $\lambda \ll 1$.
Before we turn to the problem of constructing the quasi-conserved quantities for the perturbed Hamiltonian~(\ref{H_tot}), let us briefly summarize the most important properties of the charges conserved by the integrable Hamiltonian~$H_0$.

 Local conserved charges [as those given in Eq.~(\ref{local_charges})] can be generated iteratively starting from ${\cal Q}_2 \equiv H_0$ [by convention, ${\cal Q}_1$ is the total magnetization] and using the following relation~\cite{grabowski1, grabowski2, grabowski3}:
\be \label{Q_sequence}
	{\cal Q}_{n+1} = [B,{\cal Q}_n],
\ee
where $B$ is the so-called boost operator, which reads
\be \label{B_XXX}
    B = \frac{1}{2i}  \sum_{j} j \, \boldsymbol{\sigma}_j\cdot\boldsymbol{\sigma}_{j+1}.
\ee
Thus, the boost operator acts  as a ladder operator in the space of conserved charges of the model. 
Importantly, every next charge has a larger support as compared to the previous one. For the XXX model, the $n$-th charge ${\cal Q}_n$ is a sum of operators with a support up to $n$ sites. As discussed in Ref.~\cite{grabowski2}, the conserved charges with $n>3$ generated as in~Eq.(\ref{B_XXX}) contain terms from the charges with smaller $n$. For the sake of simplicity we work with a different basis $\{Q_n\}$ in which every next charge does not contain any terms that are present in the previous ones. The first two charges coincide in both bases, i.e. ${\cal Q}_k = Q_k$ for $k=1,2$.
For completeness, here we present expressions for the few higher charges in this basis:
\be \label{unperturbed_charges}
\begin{aligned}
	&Q_3 = \sum_j  \left( \boldsymbol{\sigma}_{j} \times \boldsymbol{\sigma}_{j+1} \right) \cdot \boldsymbol{\sigma}_{j+2}, \\
	&Q_4 = \sum_j \Bigl\{  \left(\left( \boldsymbol{\sigma}_{j} \times \boldsymbol{\sigma}_{j+1} \right) \times \boldsymbol{\sigma}_{j+2} \right) \cdot \boldsymbol{\sigma}_{j+3}  \\
	&  \qquad + \boldsymbol{\sigma}_{j} \cdot \boldsymbol{\sigma}_{j+2} \Bigr\},\\
	&Q_5 = \sum_j \Bigl\{  \left(\left(\left( \boldsymbol{\sigma}_{j} \times \boldsymbol{\sigma}_{j+1} \right) \times \boldsymbol{\sigma}_{j+2} \right) \times \boldsymbol{\sigma}_{j+3} \right) \cdot \boldsymbol{\sigma}_{j+4} \\
	 & \qquad + \left( \boldsymbol{\sigma}_{j} \times \boldsymbol{\sigma}_{j+2}  +  \boldsymbol{\sigma}_{j} \times \boldsymbol{\sigma}_{j+1} \right) \cdot \boldsymbol{\sigma}_{j+3} \Bigr\}, \\
 	&Q_6 =  \sum_j \Bigl\{  \left( \boldsymbol{\sigma}_j \cdot \boldsymbol{\sigma}_{j+2} + \boldsymbol{\sigma}_j \cdot \boldsymbol{\sigma}_{j+3} \right)\\
	 &+   \bigl( \left( \boldsymbol{\sigma}_{j} \times \boldsymbol{\sigma}_{j+1}  +  \boldsymbol{\sigma}_{j} \times \boldsymbol{\sigma}_{j+2} \right) \times \boldsymbol{\sigma}_{j+3} \bigr) \cdot \boldsymbol{\sigma}_{j+4}, \\
	 & + \bigl(\bigl(\left(\left( \boldsymbol{\sigma}_{j} \times \boldsymbol{\sigma}_{j+1} \right) \times \boldsymbol{\sigma}_{j+2} \right) \times \boldsymbol{\sigma}_{j+3} \bigr) 
	\times \boldsymbol{\sigma}_{j+4} \bigr) \cdot \boldsymbol{\sigma}_{j+5}\\
	 &+ \bigl( \left( \boldsymbol{\sigma}_{j} \times \boldsymbol{\sigma}_{j+1} \right) \times \boldsymbol{\sigma}_{j+2} \bigr) \cdot \boldsymbol{\sigma}_{j+4} \Bigr\}, \\
\end{aligned}
\ee
and the general form of $Q_n$ can be found in Ref.~\cite{grabowski2}.
We emphasize once again that $B$ can only generate the {\it local} charges, whereas the Hamiltonian~(\ref{H_0}) also possesses the quasi-local ones~\cite{quasilocal1}.  To our knowledge, a corresponding boost operator that can generate quasi-local charges has not been found. 

Let us now turn on the perturbation (\ref{H_1}), such that the total Hamiltonian is $H_{\lambda}$ as given by Eq.~(\ref{H_tot}), and~$\lambda \ll 1$. The quantities~$Q_n$ are no longer conserved, since they do not commute with $H_{\lambda}$. Neither they are quasi-conserved, since $\norm{ [H_{\lambda}, Q_n] } \propto\lambda$. Hence, they change significantly over times much shorter than $t_{\text{th}} \sim \lambda^{-2}$ and can not govern the dynamics in the pre-thermal phase.


\section{Quasi-conserved charges}
\label{S:quasi_charges}


We now proceed with looking for the quasi-conserved quantities that survive during the pre-thermal phase up to times $\sim \lambda^{-2}$. This simply means that we are looking for a set of operators $\tilde{Q}_n$ that satisfy
\be \label{comm_norm_scaling}
	\norm{ [ H_{\lambda}, \tilde{Q}_n ] } \propto \lambda^2
\ee
and commute with each other with the accuracy ${\cal O}(\lambda^2)$. 

First of all, the Hamiltonian $H_{\lambda}$ has the translational and $SU(2)$ symmetries, therefore we require that the quasi-conserved charges $\tilde{Q}_n$ possess these symmetries as well. 
In analogy with the integrable case, we identify the second charge with the Hamiltonian, i.e., $\tilde{Q}_2 = H_{\lambda}$ [note that $\tilde Q_1 = Q_1$, as the total magnetization is conserved by $H_{\lambda}$]. This gives us the relation
\be
	\tilde{Q}_2 = Q_2 + \lambda H_1.
\ee
Since the perturbation is weak, it is natural to expect that similar relations should hold for higher charges as well.
Therefore, we make an ansatz
\be \label{quasiQ_ansatz}
	\tilde{Q}_{n} = Q_n + \lambda \sum_{s=m}^M \delta Q_{n}^{(s)}, 
\ee
where $\delta Q_n^{(s)}$ is a local operator consisting of terms having the support on $s$ sites. The values of $m$ and $M$ will be specified below, at the moment we can only expect that $M > n$, i.e., the maximal support of~$\tilde Q_{n}$ is larger than that of~$Q_n$. This is a reasonable assumption because the perturbed Hamiltonian $H_{\lambda}$ itself has a greater support than~$H_{0}$. 
Let us emphasize that the ansatz~(\ref{quasiQ_ansatz}) fully determines the $\lambda$-dependence of the quasi-conserved charge~$\tilde Q_n$. This is simply becasue the perturbed Hamiltonian~$H_{\lambda}$ has terms at most linear in~$\lambda$, and thus keeping in~$\tilde Q_n$ any higher order terms results in the excess of precision.
Thus, taking into account~Eq.~(\ref{quasiQ_ansatz}), one can clearly see that in order to fulfil the requirement in Eq.~(\ref{comm_norm_scaling}), the commutator~$[ H_{\lambda}, \tilde{Q}_n ]$ should not  contain terms linear in~$\lambda$, i.e., the following condition must be fulfilled:
\be \label{eliminate_linear_term}
\left[ H_1, Q_n \right] + \sum_{s = m}^{M} \bigl[ H_0, \delta Q_n^{(s)} \bigr] = 0.
\ee
In this case the commutator of $\tilde Q_n$ and $H_{\lambda}$ reads
\be \label{lambda_squared_terms}
	[ H_{\lambda}, \tilde{Q}_n ] = \lambda^2 \sum_{s=m}^{M} \bigl[ H_1, \delta Q_n^{(s)} \bigr]
\ee
and Eq.~(\ref{comm_norm_scaling}) is clearly satisfied.

Let us now discuss the structure of $\delta Q_{n}^{(s)}$ from Eq.~(\ref{quasiQ_ansatz}) in more detail. First of all, translational invariance allows us to express it in the following form:
\be \label{deltaQ}
	\delta Q_{n}^{(s)} = \sum_{k=2}^{s} \sum_{ \; \boldsymbol{\ell}_{k}(s) } c_{n}\bigl( \boldsymbol{\ell}_{k}(s) \bigr)  \sum_j O_{j} \bigl( \boldsymbol{\ell}_{k} (s)\bigr),
\ee
where $c_{n}\bigl( \boldsymbol{\ell}_{k}(s) \bigr)$ are real numerical coefficients, and the local operator $O_{j} \bigl( \boldsymbol{\ell}_{k} (s)\bigr)$ has a support on $s$ sites and acts non-trivially on $k$ sites ($2 \leq k \leq s$), specified by the components~$\ell_i$ of the vector~$\boldsymbol{\ell}_{k}(s)$:
\be \label{sites}
	j + \ell_1, \; \ldots ,\; j + \ell_{k},
\ee
\\
where $\ell_i$ take values from $\{0, \ldots, s-1 \}$.  Importantly, all~$\ell_i$ are distinct and {\it not necessarily ordered}. Note that $\boldsymbol{\ell}_{k}(s)$ always has $\ell_p = 0$ and $\ell_q = s-1$ for some indices $p,q \in \{1,\ldots,k\}$.

Due to the $SU(2)$ symmetry, $O_j\bigl( \boldsymbol{\ell}_{k}(s) \bigr)$ in Eq.~(\ref{deltaQ}) can be expressed in the basis of $SU(2)$-invariant tensor products of $k$ spin-$1/2$ operators. To construct this basis we follow Refs.~\cite{grabowski1,grabowski2,grabowski3} and introduce nested cross products of the Pauli vectors, with the nesting going toward the left:
\be \label{nested_cross_product}
\begin{aligned}
	\boldsymbol{V}_j^{(1)} &= \boldsymbol{\sigma}_{j+\ell_1},\\
	\boldsymbol{V}_j^{(2)} &= \boldsymbol{\sigma}_{j+\ell_1} \times \boldsymbol{\sigma}_{j+\ell_2},\\
	\boldsymbol{V}_j^{(3)} &= \left( \boldsymbol{\sigma}_{j+\ell_1} \times \boldsymbol{\sigma}_{j+\ell_2} \right) \times \boldsymbol{\sigma}_{j+\ell_3}, \\
	\boldsymbol{V}_j^{(k)} &= \boldsymbol{V}_j^{(k-1)} \times \boldsymbol{\sigma}_{j+\ell_k}.
\end{aligned}
\ee
Second, we consider the following scalar [and hence $SU(2)$-invariant] spin polynomials:
\be \label{basis_f} 
	f_j(\ell_1 , \ldots, \ell_k ) \equiv f_j(\boldsymbol{\ell}_k) = \boldsymbol{V}_j^{(k-1)} \cdot \boldsymbol{\sigma}_{j+\ell_k}.
\ee
For instance, for the two-spin polynomial we have $f_j(0,l) = \boldsymbol{\sigma}_{j} \cdot \boldsymbol{\sigma}_{j+l}$, for the three-spin polynimial $f_j(0,l,m) = \left( \boldsymbol{\sigma}_{j} \times \boldsymbol{\sigma}_{j+l} \right) \cdot \boldsymbol{\sigma}_{j+m}$, and for the four-spin polynomial the definition yields $f_j(l,m,n,p) = \left[\left( \boldsymbol{\sigma}_{j+l} \times \boldsymbol{\sigma}_{j+m} \right) \times \boldsymbol{\sigma}_{j+n} \right] \cdot \boldsymbol{\sigma}_{j+p}$. Then the operator~$O_{j}\bigl( \boldsymbol{\ell}_{k}(s) \bigr)$ in Eq.~(\ref{deltaQ}) can be taken as
\be \label{operatorO}
\begin{aligned}
	O_{j}& \bigl( \boldsymbol{\ell}_{k}(s) \bigr) = f_j( \ell_1, \ldots, \ell_k )\\
	= \Bigl[ &\bigl( ( \boldsymbol{\sigma}_{j+\ell_1} \times \boldsymbol{\sigma}_{ j + \ell_2} ) \times \ldots \bigr) 
	 \times \boldsymbol{\sigma}_{ j + \ell_{k-1}}  \Bigr] \cdot \boldsymbol{\sigma}_{j+\ell_k},
\end{aligned}
\ee
where~$\ell_j$ are the same as those discussed after Eq.~(\ref{sites}).
Obviously, the charges~$Q_n$ of the unperturbed model~$H_0$ can also be expressed in terms of~$f_j(\ell_1 , \ldots, \ell_k)$ from~Eq.~(\ref{basis_f}). Let us note that the corrections~$\delta Q_n^{(s)}$ in~Eq.~(\ref{quasiQ_ansatz}) are allowed to have terms that are present in the unperturbed charge~$Q_n$.
Explicitly, the spin polynomial in~Eqs.~(\ref{basis_f}) and~(\ref{operatorO}) can be written as
\be \label{f_to_sigma}
	f_j( \ell_1, \ldots, \ell_k ) = \sum_{\beta_1 \ldots \beta_k} C_{\beta_1 \ldots \beta_k} \bigotimes_{p=1}^{k} \sigma_{j+\ell_p}^{\beta_{p}},
\ee
where the coefficients~$C_{\beta_1 \ldots \beta_k}$ arise from expanding the scalar product in the spin polynomial~$f_j( \ell_1, \ldots, \ell_k )$. For~$k = 2$ and~$3$, one obviously has~$C_{\beta_1 \beta_2} =1$ and $C_{\beta_1 \beta_2 \beta_3} =\varepsilon_{\beta_1\beta_2 \beta_3}$, correspondingly, where~$\varepsilon_{ijk}$ is the Levi-Civita symbol. For $k>3$ the coefficients are given by
\be \label{f_to_sigma_coeff}
\begin{aligned}
	C_{\beta_1 \ldots \beta_k} =& \sum_{\alpha_1 \ldots \alpha_{k-3}} \varepsilon_{\alpha_1 \, \beta_{k-1}\, \beta_k } \varepsilon_{\alpha_{k-3} \, \beta_{1}\, \beta_2 } \\
	&\qquad \times \prod_{q=1}^{k-4} \varepsilon_{\alpha_q \, \alpha_{q+1}\, \beta_{k-q-1} }.
\end{aligned}
\ee

Let us briefly discuss the dimensionality of the~$SU(2)$ invariant spin polynomials~$_j(\ell_1 , \ldots, \ell_k )$. Clearly, for~$k=2$ and $3$ there is only one such operator.
For $k=4$ there are three inequivalent permutations, only two of which are linearly independent. Namely, we have  
\be \label{4spin_polynomials_linear_dependence}
f_j(0,l,m,n) - f_j(0,m,l,n) + f_j(0,n,l,m) = 0,
\ee
which one can easily check by a direct calculation using Eqs.~(\ref{f_to_sigma}) and (\ref{f_to_sigma_coeff}). Thus, for~$k=4$ the basis is two-dimensional. For $k=5$ and $k=6$ the basis is $6$- and $14$-dimensional, correspondingly.

Thus, to fully understand the operator content of the quasi-conserved charges~$\tilde Q_n$, it remains to determine the support of the corrections~$\delta Q_n^{(s)}$, i.e. the range of summation over~$s$ in~Eq.~(\ref{quasiQ_ansatz}). Let us consider the the spin polynomial~$f_{i}(m_1,\ldots,m_k)$ with the support on~$s$ sites, i.e. for some~$p$ and~$q$ we have~$m_p = 0$ and~$m_q=s-1$. Then, as  shown in Appendix~\ref{A:commutator_f_f},  the commutator~$\left[ f_{j}(0,l), \, f_{i}(m_1,\ldots,m_k) \right]$ contains terms with the support up to~$s+l$. Thus, since in general the unperturbed charge~$Q_n$ may contain~$K$ terms with the support on~$\{n_1, \ldots, n_{K-1}, n\}$~sites, with $n_1 < \ldots < n$, the commutator~$[ H_1, Q_n ]$ consists of terms with the support~$\{n_1 + 2, \dots, n_{K-1}+2, n+2\}$. 
Therefore, we immediately see that in order to satisfy Eq.~(\ref{eliminate_linear_term}) the correction~$\sum_{s=m}^{M} \delta Q_n^{(s)}$ must include terms having the support on~$\{n_1 + 1, \ldots, n_{K-1}+1, n+1\}$ sites, such that the maximal support  of the commutator~$[H_0, \sum_{s=m}^{M} \delta Q_n^{(s)}]$ matches that of~$[ H_1, Q_n ]$ and the two commutators cancel each other. Therefore, for the quasi-conserved charges we finally have 
\be \label{quasiQ_ansatz_2}
	\tilde{Q}_{n} = Q_n + \lambda \sum_{s=n_1}^{n+1} \delta Q_{n}^{(s)}, 
\ee
where~$n_1$ and $n$ are, correspondingly, the smallest and the largest support of the terms in~$Q_n$. Note that some of the terms in~Eq.~(\ref{quasiQ_ansatz_2}) may have zero coefficient. For instance, as shown in the next Section, in~$\tilde Q_3$ the only corrections that are present have the support~$s =n+1 = 4$. With this, we now proceed to investigating the possibility of satisfying~Eqs.~(\ref{comm_norm_scaling}) and ~(\ref{eliminate_linear_term}).

\section{Results and Discussion}
\label{S:results}

Taking into account Eqs.~(\ref{quasiQ_ansatz}), (\ref{deltaQ}), (\ref{basis_f}), and (\ref{operatorO}) to construct the ansatz for the quasi-conserved charges~$\tilde Q_n$, we fix the coefficients~$c_{n}\bigl( \boldsymbol{\ell}_{k}(s) \bigr)$ in~Eq.~(\ref{deltaQ}) such that the criterion~(\ref{eliminate_linear_term}) is satisfied, which  guarantees the required scaling behavior of the commutator norm in Eq.~(\ref{comm_norm_scaling}). In order to evaluate the commutators in~Eq.~(\ref{eliminate_linear_term}) we used the results of Appendix~\ref{A:commutators}.
Then, for the lowest order quasi-conserved charge~$\tilde Q_3$ we obtain
\be \label{tildeQ3}
\begin{aligned}
	& \tilde Q_3 = Q_3 + \lambda \, \delta Q_3^{(4)}, \\
	& \delta Q_3^{(4)} = \sum_j  \left( \boldsymbol{\sigma}_{j} \times \boldsymbol{\sigma}_{j+1} + \boldsymbol{\sigma}_{j} \times \boldsymbol{\sigma}_{j+2} \right) \cdot \boldsymbol{\sigma}_{j+3}.
\end{aligned}
\ee
From Eqs.~(\ref{unperturbed_charges}) and~(\ref{tildeQ3}) we see that the correction~$\delta Q_3^{(4)}$ has a structure resembling that of~$Q_3$, i.e., every term in $\delta Q_3^{(4)}$ has the same number of spins, but the support is increased by one site. Let us now look at the norm of~$[\tilde Q_3, H_{\lambda}]$. For computational convenience we use Frobenius norm defined as
\be \label{Frobenius_norm_def}
	\norm{X}_F = \sqrt{ \text{tr}\left( X^{\dag} X \right) }.
\ee
Then, from~Eqs.~(\ref{lambda_squared_terms}) and~(\ref{Frobenius_norm_def}) we have
\be \label{commutator_norm_Q3}
	\frac{1}{\norm{H_0}_F}\norm{\bigl[ H_{\lambda}, \tilde Q_3 \bigr]}_F = 8\, \lambda^2,
\ee
where~$\norm{H_0}_F = \sqrt{3 \, N} \,2^{N/2}$ is introduced for normalization.

The situation with the higher order charges is slightly more involved. It turns out that one can find families of quasi-conserved quantities~$\tilde Q_n$ satisfy~Eq.~(\ref{eliminate_linear_term}) with $n>3$. In other words, the higher order quasi-conserved charges $\tilde Q_n$ contain a number of free parameters.
 Repeating the steps from the beginning of this~Section, we obtained the families of quasi-conserved charges~$\tilde Q_n$ with $n = 4$, $5$, and $6$. The latter two, $\tilde Q_5$ and $\tilde Q_6$, contain a large number of terms and for the sake of readability, we present their explicit form in~Appendix~\ref{A:quasi-charges}. Here we only give the expression for the family of~$\tilde Q_4$, which reads
\be \label{tildeQ_4}
\begin{aligned}
	& \tilde Q_4 = Q_4 +\lambda \sum_{s=3}^{5} \delta Q_4^{(s)}, \\
	& \delta Q_4^{(3)} = (a - 2 ) \sum_{j}\, \boldsymbol{\sigma}_j\cdot\boldsymbol{\sigma}_{j+2}, \\
	& \delta Q_4^{(4)} = \sum_{j}\, \Bigl\{ \boldsymbol{\sigma}_j\cdot\boldsymbol{\sigma}_{j+3} \\
	& \qquad + a  \left(\left( \boldsymbol{\sigma}_{j} \times \boldsymbol{\sigma}_{j+1} \right) \times \boldsymbol{\sigma}_{j+2} \right) \cdot \boldsymbol{\sigma}_{j+3},\\
	& \qquad -  \left(\left( \boldsymbol{\sigma}_{j} \times \boldsymbol{\sigma}_{j+2} \right) \times \boldsymbol{\sigma}_{j+1} \right) \cdot \boldsymbol{\sigma}_{j+3} \Bigr\},\\
	& \delta Q_4^{(5)} =  \sum_j \Bigl\{  \left(\left( \boldsymbol{\sigma}_{j} \times \boldsymbol{\sigma}_{j+1} \right) \times \boldsymbol{\sigma}_{j+2} \right) \cdot \boldsymbol{\sigma}_{j+4}\\
	&\qquad +   \left(\left( \boldsymbol{\sigma}_{j} \times \boldsymbol{\sigma}_{j+1} \right) \times \boldsymbol{\sigma}_{j+3} \right) \cdot \boldsymbol{\sigma}_{j+4} \\
	& \qquad +   \left(\left( \boldsymbol{\sigma}_{j} \times \boldsymbol{\sigma}_{j+2} \right) \times \boldsymbol{\sigma}_{j+3} \right) \cdot \boldsymbol{\sigma}_{j+4} \Bigr\}. \\
\end{aligned}
\ee
We stress that~Eq.~(\ref{eliminate_linear_term}) is satisfied for arbitrary~$a$. Thus, one can use this freedom and minimize the number of terms in~$\tilde Q_4$. This is achieved for~$a = 0$ or $a = 2$. On the other hand, the coefficient $a$ in Eq.~(\ref{tildeQ_4}) can be used as a variational parameter to further minimize the norm of~$[\tilde Q_4, H_{\lambda}]$, which is given by
\be \label{commutator_norm_Q4}
	\frac{1}{\norm{H_0}_F}\norm{\bigl[ H_{\lambda}, \tilde Q_4 \bigr]}_F = 4\sqrt{2}\, \lambda^2 \, \bigl( 4a^2 + a +10 \bigr)^{1/2}.
\ee
Eq.~(\ref{commutator_norm_Q4}) is minimal at~$a = - 1/8$ and equals $\sqrt{318} \lambda^2$, whereas for~$a = 0$ the squared  norm~(\ref{commutator_norm_Q4}) becomes~$\sqrt{320} \lambda^2$. The difference of~approximately $0.6\%$ is negligibly small for our purposes, and 
one may use a simpler form of~$\tilde Q_4$ with~$a = 0$. Similar analysis for the next quasi-conserved charges~$\tilde Q_5$ and~$\tilde Q_6$ can be found in Appendix~\ref{A:quasi-charges}. We also point out that by looking at the (\ref{commutator_norm_Q4}) (as well as at the expressions from higher quasi-conserved charges) it would be tempting to suggest that if $a$ is such that is solves the quadratic equation $4a^2 + a +10=0$, the quasi-conserved charge becomes exactly conserved. This would require $a$ to be complex. We checked that the whole procedure should be modified then and the final answer is that the overlap always stays finite, even for complex perturbations.  

One can check by a lengthy but straightforward direct calculation that the commutator of the quasi-conserved charges $[ \tilde{Q}_n, \tilde{Q}_m ]$ does not contain terms linear in $\lambda$. We expect that this is correct for any~$n$ and $m$ and have checked this explicitly for~$\tilde Q_n$ with $3\geq n \geq 6$. It is also straightforward to see that any linear combination of the quasi-conserved charges is a quasi-conserved charges itself. 


To summarize, in this paper we have shown that an isotropic Heisenberg spin chain, weakly perturbed away from integrability by a next to nearest neighbour interaction of strength $\lambda$, possesses quasi-conserved charges $\tilde Q_n$ with~$3\leq n \leq 6$, which are approximately conserved up to times of the order $\lambda^{-2}$. We conjecture that the perturbed model~$H_{\lambda}$ from Eq.~(\ref{H_tot}) has as many quasi-conserved charges~$\tilde Q_n$ as there are conserved charges~$Q_n$ for the integrable model~(\ref{H_0}), but the proof of our conjecture is beyond the scope of the present paper. We also expect that our results can be extended to the case of other perturbations and open boundary conditions, as well as to other one-dimensional models with~$SU(2)$ symmetry, e.g. the  anisotropic Heisenberg chains (XXZ and XYZ models) and the Hubbard model, which we leave for future studies. Presence of these quasi-conserved charges could affect some transport properties, see e.g.\cite{Rosch2006}, \cite{Prosen2015}.  

\section{Acknowledgements}
We would like to thank Anatoli Polkovnikov for very useful discussions leading to this work. We also thank Balasz Pozsgay for drawing our attention to Refs.~\cite{longrange1, longrange2, Pozsgay2020, Marchetto2020} and Anatoli Dymarsky for very useful comments. This work is part of the DeltaITP consortium, a program of the Netherlands Organization for Scientific Research (NWO) that is funded by
the Dutch Ministry of Education, Culture and Science
(OCW).
The results of D.V.K. were supported by the Russian Science
Foundation Grant No. 20-42-05002 (parts of Secs.~\ref{S:model_boost}, \ref{S:quasi_charges}, and~\ref{S:results}).

\appendix
\pagebreak 
\begin{widetext}
\section{} \label{A:commutator_f_f}
Let us consider the commutator of $f_{i}(0,l)$ and $f_{j}(m_1,\ldots,m_k)$:


\be \label{commutator_f_f}
\begin{aligned}
	&\Bigl[ f_{j}(0,l), \, f_{i}(m_1,\ldots,m_k) \Bigr] 
	&= \sum_{i, j} \sum_{\alpha}\sum_{\beta_1 \ldots \beta_k} C_{ \beta_1 \ldots \beta_k } \Bigl[ \sigma_{j}^{\alpha} \sigma_{{j}+l}^{\alpha} \, , \bigotimes_{p=1}^{k} \sigma_{i+m_p}^{\beta_{p}} \Bigr], 
\end{aligned}
\ee
where the polynomial~$f_{i}(m_1,\ldots,m_k)$ contains terms with~$k$ Pauli matrices and is assumed to have the support on $s$~sites. Then, we rewrite the commutator on the right hand side of Eq.~(\ref{commutator_f_f}) as
\be
	\sum_{i,j}\Bigl[ \sigma_{i}^{\alpha} \sigma_{{i}+l}^{\alpha} \, , \bigotimes_{p=1}^{k} \sigma_{j+m_p}^{\beta_{p}} \Bigr] = \sum_{j} \sum_{p=1}^k \Bigl( {\cal A}_j(p) + (1-\delta_{l,m_p}) \prod_{\substack{q=1 \\ q \neq p}}^k \left( 1- \delta_{l,m_p - m_q}  \right) {\cal B}_j(p) \Bigr),
\ee
where ${\cal A}_j(p)$ and ${\cal B}_j(p)$ arise from the terms in~Eq. (\ref{commutator_f_f}) with~$i = j+m_p$ and~$i+l = j+m_p$, correspondingly. Explicitly, we have
\be
	{\cal A}_j(p) = \Bigl[ \sigma_{j}^{\alpha} \sigma_{{j}+l}^{\alpha} \, , \sigma_j^{\beta_p}\bigotimes_{\substack{ q=1 \\ q\neq p }}^{k} \sigma_{j + m_q - m_p}^{\beta_{q}} \Bigr], \qquad
	{\cal B}_j(p) = \sigma_{{j}}^{\alpha}  \Bigl[ \sigma_{j+l}^{\alpha} \, , \sigma_{j+l}^{\beta_p} \Bigr] \bigotimes_{\substack{ r=1 \\ r\neq p }}^{k} \sigma_{j +l + m_r - m_p}^{\beta_{r}}.
\ee
It is straightforward to see that
\be
	{\cal A}_j(p) = \sum_{\substack{q=1 \\ q \neq p}}^k \delta_{l,m_q-m_p} {\cal A}_j^{(1)}(p,q) +\prod_{\substack{q=1 \\ q \neq p}}^k \left(1 -\delta_{l,m_q-m_p} \right) {\cal A}_j^{(2)}(p) ,
\ee
where
\be
	{\cal A}_j^{(1)}(p,q) = \Bigl[ \sigma_{j}^{\alpha} \sigma_{{j}+l}^{\alpha} \, , \sigma_j^{\beta_p} \sigma_{j+l}^{\beta_q}\Bigr] \bigotimes_{\substack{ r=1 \\ r\neq p, q }}^{k} \sigma_{j + m_r - m_p}^{\beta_{r}}, \qquad 
	{\cal A}_j^{(2)}(p) = \Bigl[ \sigma_{j}^{\alpha} \, , \sigma_j^{\beta_p} \Bigr] \sigma_{{j}+l}^{\alpha} \bigotimes_{\substack{ r=1 \\ r\neq p }}^{k} \sigma_{j + m_r - m_p}^{\beta_{r}}.
\ee
Thus, we obtian
\be \label{A_1_2_B}
\begin{aligned}
	{\cal A}_j^{(1)}(p,q) &= 2i \Bigl( \delta_{\alpha, \beta_q} \sum_{\gamma_p} \varepsilon_{\alpha,\beta_p, \gamma_p} \sigma_j^{\gamma_p} 
	+ \delta_{\alpha, \beta_p} \sum_{\gamma_q} \varepsilon_{\alpha,\beta_q, \gamma_q} \sigma_{j+l}^{\gamma_q} \Bigr)\bigotimes_{\substack{ r=1 \\ r\neq p, q }}^{k} \sigma_{j + m_r - m_p}^{\beta_{r}},\\
	{\cal A}_j^{(2)}(p) &= 2i \sum_{\gamma_p} \varepsilon_{\alpha,\beta_p, \gamma_p} \sigma_j^{\gamma_p}  \sigma_{{j}+l}^{\alpha} \bigotimes_{\substack{ r=1 \\ r\neq p}}^{k} \sigma_{j + m_r - m_p}^{\beta_{r}}, \qquad 	{\cal B}_j(p) = 2i \, \sigma_{{j}}^{\alpha} \, \sum_{\gamma_p} \varepsilon_{\alpha,\beta_p, \gamma_p} \sigma_{j+l}^{\gamma_p}   \bigotimes_{\substack{ r=1 \\ r\neq p }}^{k} \sigma_{j + l + m_r - m_p}^{\beta_{r}}.
\end{aligned}
\ee

It is now easy to see that the commutator in Eq.~(\ref{commutator_f_f}) consists of terms with the support not greater than~$s+l$. Indeed, let us consider, e.g.~${\cal B}_j(p)$ from Eq.~(\ref{A_1_2_B}) and take~$m_p = 0$. Then, for some~$r=r_*$ me have $m_{r_*} = s-1$ and the Pauli matrix with the largest index is~$\sigma_{j+l+s-1}^{\beta_{r_*}}$, whereas the Pauli matrix with the smallest index is~$\sigma_j^{\alpha}$. Clearly, such term has a support on $s+l$ sites.

\section{} \label{A:commutators}
We use the following identities
\be
\begin{aligned}
	& \left[ \boldsymbol{A} \cdot \boldsymbol{X}, \boldsymbol{A} \cdot \boldsymbol{Y} \right] = 2i \left( \boldsymbol{X} \times \boldsymbol{Y}\right) \cdot \boldsymbol{A},\\
	& \left[ \boldsymbol{A} \cdot \boldsymbol{B}, (\boldsymbol{B}\times \boldsymbol{L})\cdot \boldsymbol{R} \right] = -2i \left( \left(\boldsymbol{A} \times \boldsymbol{B} \right) \times \boldsymbol{L} \right) \cdot \boldsymbol{R}, \\
	& \left[ \boldsymbol{A} \cdot \boldsymbol{B}, (\boldsymbol{A}\times \boldsymbol{B})\cdot \boldsymbol{R} \right] = 4i \left( \boldsymbol{B} \cdot \boldsymbol{R} - \boldsymbol{A} \cdot \boldsymbol{R} \right), \\
	&  \left[ \boldsymbol{A} \cdot \boldsymbol{B}, \left( (\boldsymbol{L}\times \boldsymbol{A}) \times \boldsymbol{B} \right)\cdot \boldsymbol{R} \right] = 2i \left( \left(\boldsymbol{L} \times \boldsymbol{B} \right) \cdot \boldsymbol{R} -\left(\boldsymbol{L} \times \boldsymbol{A} \right) \cdot \boldsymbol{R} \right),
\end{aligned}
\ee
where $\boldsymbol{A} = \{ \sigma_{j_1}^{x}, \sigma_{j_1}^{y}, \sigma_{j_1}^{z} \}$, $\boldsymbol{B} = \{ \sigma_{j_2}^{x}, \sigma_{j_2}^{y}, \sigma_{j_2}^{z} \}$, and~$\boldsymbol{L}$, $\boldsymbol{R}$ are arbitrary tensor products that commute with both~$\boldsymbol{A}$ and $\boldsymbol{B}$. Then, one can show that
\be
	\sum_{j,k} \left[ f_j(0,l), f_k(0,m)  \right] = 2i \sum_j \Bigl( f_j(0,l,m) + f_j(0,m,l+m) - f_j(0,l,l+m) -f_j(0,m-l,m) \Bigr), \qquad l \neq m,
\ee
and zero if $l = m$.
\be \label{f_comm_2_3}
\begin{aligned}
	\sum_{j,k}& \left[ f_j(0,l), f_k(0,m,n)  \right] = 2 i \sum_j \Bigl\{  (1-\delta _{l,m}-\delta _{l,n}) f_j(0,l,m,n) + 2 \delta _{l,m} \left[ f_j(0,n-l)-f_j(0,n)\right]  \\
	&- 2 \delta _{l,n} [ f_j(0,m-l)-f_j(0,m)] - (1-\delta _{l,n-m}) f_j(m,l+m,0,n) + 2 \delta _{l,n-m} [ f_j(0,l+m)-f_j(0,m) ] \\
	& +  f_j (n,l+n,0,m) - f_j (0,l,l+m,l+n) +  (1-\delta _{l,m}) f_j (0,l,l-m,l-m+n)\\
	& - (1-\delta _{l,n}) (1-\delta _{l,n-m}) f_j(0,l,l-n,l+m-n)  \Bigr\},
\end{aligned}
\ee
and 
\be \label{f_comm_2_4}
\begin{aligned}
	\sum_{j,k}& \left[ f_j(0,l), f_k(m,n,p,q)  \right] = 2 i \sum_j \Bigl\{ K(l, m, n, p, q) - K(l, n, m, p, q) + K(l, p, q, m, n ) - K(l, q, p, m, n)\Bigr\},
\end{aligned}
\ee
where we denoted
\be \label{f_comm_2_4_K}
\begin{aligned}
	K(l, m, n, p, q) &= 4 \delta _{l,n-m} [ f(0,-l-m+p,-l-m+q)-f(0,p-m,q-m) ] \\
	&-2 \delta _{l,p-m} [f(-l-m+n,0,-l-m+q)-f(n-m,0,q-m) ]\\
	&-2 \delta _{l,q-m} [f(-l-m+p,0,-l-m+n)-f(p-m,0,n-m)]  \\
	& + 2 (1 - \delta _{l,n-m}-\delta _{l,p-m}-\delta _{l,q-m}) f(0,l,n-m,p-m,q-m) \\
	& -2 (1-\delta _{l,m}) (1-\delta _{l,m-n}) (1-\delta _{l,m-p}) (1-\delta _{l,m-q})
   f(0,l,l-m+n,l-m+p,l-m+q).
\end{aligned}
\ee

\section{} \label{A:quasi-charges}

The basis of $5$-spin polynomial is $6$ dimensional and one has the following relations:
\be
	\begin{pmatrix}
		f_j (l, q, m, n, p) \\
		f_j(l, p, q, m, n) \\
		f_j(l, p, m, n, q) \\
		f_j(l, n, q, m, p) \\
		f_j(l, n, p, m, q) \\
		f_j(l, n, m, p, q) \\
		f_j(l, m, q, n, p) \\
		f_j(l, m, p, n, q) \\
		f_j(l, m, n, p, q) \\
	\end{pmatrix} = 
	   \begin{pmatrix} 
	       0 & 1 & -1 & 0 & 0 & 0 \\
	       0 & 0 & 1 & 1 & 0 & 0 \\
	       1 & 0 & -1 & -1 & 0 & 0 \\
	       0 & 1 & 0 & 0 & 1 & 0 \\
	       1 & 0 & 0 & 0 & 0 & 1 \\
	       1 & -1 & 0 & 0 & -1 & 1 \\
	       0 & 1 & -1 & 0 & 0 & -1 \\
	       1 & 0 & -1 & -1 & -1 & 0 \\
	       1 & -1 & 0 & -1 & -1 & 1 \\
	   \end{pmatrix}
	   \begin{pmatrix}
		f_j (l, p, n, m, q) \\
		f_j(l, q, n, m, p) \\
		f_j(l, q, p, m, n) \\
		f_j(m, n, l, p, q) \\
		f_j(m, p, l, n, q) \\
		f_j(m, q, l, n, p) \\
	\end{pmatrix}.
\ee
For the quasi-conserved charge~$\tilde Q_5$ we have
\be \label{tildeQ5}
\begin{aligned}
	& \tilde Q_5 = Q_5 +\lambda \sum_{s=4}^{6} \delta Q_5^{(s)}, \\
	& \delta Q_5^{(4)} = - a \, \sum_j \Bigl[ f_j(0,1,3) + f_j(0,2,3) \Bigr], \\
	& \delta Q_5^{(5)} = \sum_j \Bigl[ f_j(0,1,4) + 2 f_j(0,2,4) + f_j(0,3,4) + a f_j(0,4,2,1,3) + a f_j(1,2,0,3,4) + \left(1 + a \right)  f_j(1,3,0,2,4) \Bigr] \\
   	& \qquad\quad + \sum_j \Bigl[ f_j(0,4,3,1,2) - a f_j(1,4,0,2,3) -\left( 1 + a \right) f_j(0,3,2,1,4)  \Bigr],\\
   	& \delta Q_5^{(6)} = \sum_j \Bigl[  f_j(0,3,2,1,5) - f_j(0,5,2,1,3) - f_j(1,2,0,3,5) - f_j(1,3,0,2,5) + f_j(1,5,0,2,3) \Bigr] \\
	&\qquad\quad+\sum_j \Bigl[ f_j(0,4,2,1,5) - f_j(0,5,2,1,4) - f_j(1,2,0,4,5) - f_j(1,4,0,2,5) + f_j(1,5,0,2,4) \Bigr] \\
	&\qquad\quad+\sum_j \Bigl[ f(_j0,4,3,1,5) - f_j(0,5,3,1,4) - f_j(1,3,0,4,5) - f_j(1,4,0,3,5) + f_j(1,5,0,3,4) \Bigr] \\
	&\qquad\quad+\sum_j \Bigl[ f_j(0,4,3,2,5) - f_j(0,5,3,2,4) - f_j(2,3,0,4,5) - f_j(2,4,0,3,5) + f_j(2,5,0,3,4) \Bigr]. \\
\end{aligned}
\ee
Using Eq.~(\ref{Frobenius_norm_def}), for the norm of~$[H_{\lambda}, \tilde Q_{5} ]$ we obtain
\be \label{commutator_norm_Q5}
	\frac{1}{\norm{H_0}_F}\norm{\bigl[ H_{\lambda}, \tilde Q_5 \bigr]}_F = 8\, \lambda^2 \, \bigl( 5a^2 
	 - 7a +20 \bigr)^{1/2}.
\ee
At~$a = 0.7$ the norm is minimal and the right hand side of~Eq.~(\ref{commutator_norm_Q5}) becomes $\approx 33.5 \lambda^2$, whereas at~$a = 0$ it is equal~$16\sqrt{5} \lambda^2 \approx 35.8 \lambda^2 $ and the difference is around~$6 \%$.

The basis of~$6$-spin polynomials is $14$-dimensional and it can be chosen as
\be
\begin{aligned}
\Bigl\{ & f_j(2,4,0,3,1,5), f_j(2,4,1,0,3,5), f_j(2,4,1,3,0,5), f_j(2,4,3,0,1,5), f_j(2,4,3,1,0,5), \\
   &f_j(3,4,0,1,2,5), f_j(3,4,0,2,1,5), f_j(3,4,1,0,2,5), f_j(3,4,1,2,0,5), f_j(3,4,2,0,1,5), \\
   & f_j(3,4,2,1,0,5), f_j(1,4,3,2,0,5), f_j(1,4,3,0,2,5), f_j(0,4,3,2,1,5)\Bigr\}.
\end{aligned}
\ee
The quasi-conserved charge~$\tilde Q_6$ has the following form:
\be \label{tildeQ6_1}
\begin{aligned}
	\tilde Q_6 &= Q_6 + \lambda \sum_{s=3}^{7} \delta Q_6^{(s)}, \\
	\delta Q_6^{(3)} &= 2 (a - 1) \sum_j  f_j(0, 2), \qquad \delta Q_6^{(4)} = (2 a + 3) \sum_j f_j (0, 3), \\
	\delta Q_6^{(5)} &= \sum_j \Bigl\{ f_j (0, 4) + 2 a f_j (0,2,1,4) + 2(a + 1) f_j(0,3,1,4)+2 a f_j(0,3,2,4) - (1+ 2 a) f_j (0,4,1,2) \\
	& \qquad\qquad -2 (1+ a) f_j (0,4,1,3)  - (1+ 2 a) f_j (0,4,2,3) \Bigr\}, \\
	\delta Q_6^{(6)} &= \sum_j \Bigl\{ f_j (0,2,1,5) + 2 f_j (0,3,1,5) + 2 f_j (0,3,2,5) + f_j (0,4,1,5) + 2 f_j (0,4,2,5) + f(0,4,3,5) \\
	&\qquad - f_j (0,5,1,2) - 2 f_j (0,5,1,3) - f_j (0,5,1,4) - 2 f_j (0,5,2,3) - 2 f_j (0,5,2,4) - f_j (0,5,3,4) \\
	& + (2 a - 1) f_j (0,4,3,2,1,5) + (1 - 2 a) f_j (1,4,3,2,0,5) - a f_j (3,4,0,1,2,5) + (1 - a) f_j (3,4,0,2,1,5) \\
	&+ a f_j (3,4,1,0,2,5) + a f_j (3,4,1,2,0,5) + a f_j (3,4,2,0,1,5) - (1 + a) f_j (3,4,2,1,0,5) \Bigr\}, \\
\end{aligned} 
\ee
\be \label{tildeQ6_2}
\begin{aligned}
	\delta Q_6^{(7)} &= \sum_j \Bigl\{ f_j (0,4,3,2,1,6) + f_j (0,5,3,2,1,6) + f_j (0,5,4,2,1,6) + f_j (0,5,4,3,1,6) + f_j (0,5,4,3, 2, 6) \\
	& - f_j (1,4,3,0,2,5) - f_j (1,4,3,2,0,6) - f_j (1,5,3,2,0,6) - f_j (1,5,4,2,0,6) - f_j (1,5,4,3,0,6) \\
	& + f_j (2,4,0,3,1,5) - f_j (2,4,1,3,0,5) + f_j (2,4,3,1,0,5) - f(2,5,4,3,0,6) \\
   &-\frac{1}{2} f_j (3,4,0,1,2,6) - \frac{1}{2} f_j (3,4,0,2,1,6)+\frac{1}{2} f_j (3,4,1,0,2,6)
   +\frac{1}{2} f_j (3,4,1,2,0,6) + \frac{1}{2} f_j (3,4,2,0,1,6) \\
   & - \frac{1}{2} f_j (3,4,2,1,0,6) - \frac{1}{2}
   f_j (3,5,0,1,2,6) - \frac{1}{2} f_j (3,5,0,2,1,6) + \frac{1}{2} f_j (3,5,1,0,2,6) 
   +\frac{1}{2} f(3,5,1,2,0,6) \\
   &+\frac{1}{2} f(3,5,2,0,1,6)-\frac{1}{2} f(3,5,2,1,0,6)-\frac{1}{2}
   f(4,5,0,1,2,6)-\frac{1}{2} f(4,5,0,1,3,6)-\frac{1}{2} f(4,5,0,2,1,6) \\
   & -\frac{1}{2}
   f(4,5,0,2,3,6)-\frac{1}{2} f(4,5,0,3,1,6)-\frac{1}{2} f(4,5,0,3,2,6)+\frac{1}{2}
   f(4,5,1,0,2,6)+\frac{1}{2} f(4,5,1,0,3,6) \\
   & +\frac{1}{2} f(4,5,1,2,0,6)+\frac{1}{2}
   f(4,5,1,3,0,6)+\frac{1}{2} f(4,5,2,0,1,6)+\frac{1}{2} f(4,5,2,0,3,6)-\frac{1}{2}
   f(4,5,2,1,0,6) \\ 
   & +\frac{1}{2} f(4,5,2,3,0,6)+\frac{1}{2} f(4,5,3,0,1,6)+\frac{1}{2}
   f(4,5,3,0,2,6)-\frac{1}{2} f(4,5,3,1,0,6)-\frac{1}{2} f(4,5,3,2,0,6) \Bigr\}.
\end{aligned} 
\ee
Then, using Eq.~(\ref{Frobenius_norm_def}), for the commutator norm we obtain
\be \label{commutator_norm_Q6}
	\frac{1}{\norm{H_0}_F}\norm{\bigl[ H_{\lambda}, \tilde Q_6 \bigr]}_F = 8\, \lambda^2 \, \bigl( 66a^2 
	 + 102 a + 113 \bigr)^{1/2},
\ee
at $a=-17/22 \approx 0.77$ we have $\approx 68.6 \lambda^2$, at $a = 0$ we have $\approx 85 \lambda^2$.
\end{widetext}

\end{document}